# ESTUDO DA INSTABILIDADE KELVIN-HELMHOLTZ ATRAVÉS DE SIMULAÇÕES COM O CÓDIGO ATHENA

# STUDY OF THE KELVIN-HELMHOLTZ INSTABILITY THROUGH SIMULATIONS WITH THE CODE ATHENA

Priscila Freitas-Lemes[1]
Irapuan Rodrigues[2]
Max Faúndez-Abans[3]

**RESUMO:** *As instabilidades Kelvin-Helmholtz são comuns em sistemas astrofísicos e vão desde jatos de buracos negros até disco de acreção protoplantário. Um objeto astrofísico com fortes características da instabilidade de Kelvin-Helmholtz é a Nebulosa de Caraguejo, na qual a expansão do material foi ocasionado pela explosão de uma supernova há, aproximadamente, 1000 anos. Essa instabilidade ocorre no limite entre dois fluidos de diferentes densidades, quando um dos fluidos é acelerado com relação ao outro. Com o objetivo de estudar essa instabilidade, realizamos uma simulação com o código de malha euleriana ATHENA. Para essa simulação, consideramos um domínio quadrado com limites periódicos sobre as laterais, e, refletindo na fronteira da parte superior e inferior. A região superior da caixa é preenchida com um gás de densidade ρ=1,0, pressão P1=1,0, índice adiabático γ=5/3 e velocidade u1=0,03 na direção x (para direita). A parte inferior tem densidade ρ=2,0, mesma pressão, velocidade e índice adiabático, só que no sentido contrário, para a esquerda. A velocidade é definida como uma função senoidal, que cria a perturbação inicial. Como resultado, observamos o princípio da instabilidade e a formação dos vórtices, com as cristas bem definidas. A nitidez da fronteira entre o material de alta e de baixa densidade está bem conservada, devido à difusão relativamente baixa do algoritmo. Notamos, ainda, que, evoluindo a simulação, os vórtices formados a partir da turbulência fundem-se.*
**Palavras-Chave:** Kelvin-Helmholt; simulações hidrodinâmicas.

**ABSTRACT:** *Kelvin-Helmholtz instabilities are common in astrophysical systems, ranging from jet black holes up to protoplanetary accretion disk. An astrophysical object with strong characteristics of the Kelvin-Helmholtz instability is Caraguejo Nebula, in which the material expansion was caused by the explosion of a supernova about 1000 years ago. This instability occurs at the boundary between two fluids of different densities when one of the fluids accelerated with respect to the other. In order to study this instability, we performed a simulation with the code ATHENA Eulerian mesh. For this simulation, we consider a square domain with periodic boundaries on the sides, and reflecting on the boundary of the top and bottom. The upper box is filled with a gas density ρ = 1.0, pressure P1 = 1.0, adiabatic index γ = 5/3, and velocity u1 = 0.03 in the x direction (to the right). The lower portion has a density ρ = 2.0, the same pressure, velocity, and adiabatic index, only in the opposite direction to the left. Speed is defined as a sinusoidal function, which creates the initial disturbance. As a result, we observe the principle of instability and the formation of vortices, with well-defined ridges. The distinctness of the boundary between the material of high and low density is well preserved due to the relatively low diffusion algorithm. We also note that the simulation evolving vortices formed from the turmoil merge.*

[1] Doutoranda em Física e Astronomia – Universidade do Vale do Paraíba - UNIVAP / Instituto de Pesquisa e Desenvolvimento - IP&D. E-mail: priscila@univap.br.
[2] Docente da Univap / IP&D. E-mail: irapuan@univap.br.
[3] Ministério da Ciência, Tecnologia e Inovação - MCTI / Laboratório Nacional de Astrofísica - LNA. E-mail: max@lna.br.

## 1. INTRODUÇÃO

A instabilidade de Kelvin-Helmholtz (K-H) é um fenômeno que surge em camadas de cisalhamento, isto é, onde se enfrentam duas regiões com velocidades distintas, ou quando houver suficiente diferença de velocidade ao longo de uma região de fricção entre dois fluidos (NAPPO, 2002).

Esse fenômeno é uma das mais famosas e onipresentes instabilidades da mecânica dos fluidos (MCLACHLAN; MARSLAND, 2006). Estudada inicialmente por Helmholtz, em 1868, e analisada, matematicamente, por Kelvin, 1871, essa instabilidade é caracterizada por uma camada de vorticidade que se enrola formando as estruturas características de Kelvin-Helmholtz.

A Figura 1 ilustra a ocorrência dessa instabilidade em três situações distintas. Na primeira, são vistas as ondas de K-H nas nuvens. A segunda figura mostra a instabilidade K-H, nas imagens, do ultravioleta extremo das ejeções de massa coronal (FOULLON et al., 2013). A terceira e última imagem mostra a instabilidade de K-H, capturada pela sonda espacial Cassini-Huygens na atmosfera de Saturno.

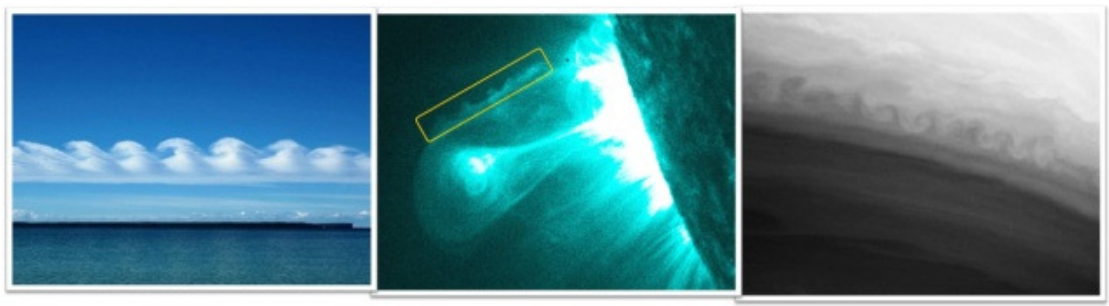

**Figura 1 - (a) Instabilidades de K-H visíveis em nuvens; (b) Ondas características da instabilidade K-H na atmosfera solar (FOULLON et al., 2011); (c) Imagem da instabilidade K-H capturada pela sonda Cassini, em outubro de 2004, na atmosfera de Saturno.**

A instabilidade de Kelvin-Helmholtz é uma das instabilidades fundamentais de fluidos incompressíveis e ocorre em um escoamento cisalhante entre dois fluidos imiscíveis. O movimento da superfície livre que separa os dois fluidos cisalhantes é controlado, dinamicamente, pelos efeitos da tensão superficial e da viscosidade.

Visto que essa instabilidade está presente em diversos objetos da astrofísica, este trabalho tem como objetivo simular a instabilidade de K-H usando o código de malha euleriana ATHENA.

## 2. O CÓDIGO ATHENA

O código Athena foi desenvolvimento na Universidade de Maryland, liderada por J. Stone, T. Gardiner e P. Teuben. O objetivo do código era o estudo do meio interestelar, formação de estrelas e fluxos de acreção (STONE, 2000). Hoje, a utilização desse código vai desde testes de convergência de disco de acréscimo (HAWLEY et al., 2013) até estudos sobre a compreensão dos impactos da rotação nas estruturas e composição de estrelas de nêutrons (WEBER; ORSARIA; NEGREIROS, 2012).

## 3. A SIMULAÇÃO

Neste estudo da instabilidade de K-H usando o código Athena, nós consideramos um domínio quadrado com limites periódicos sobre as laterais, e, refletindo na fronteira da parte superior e inferior. A região superior da caixa é preenchida com um gás de densidade ρ=1,0, pressão P1=1,0, índice adiabático γ=5/3 e velocidade u1=0,03 na direção x (para direita). A parte inferior tem densidade ρ=2,0, mesma pressão (P2=P1) e mesma velocidade e índice adiabático, só que no sentido contrário, para a esquerda.

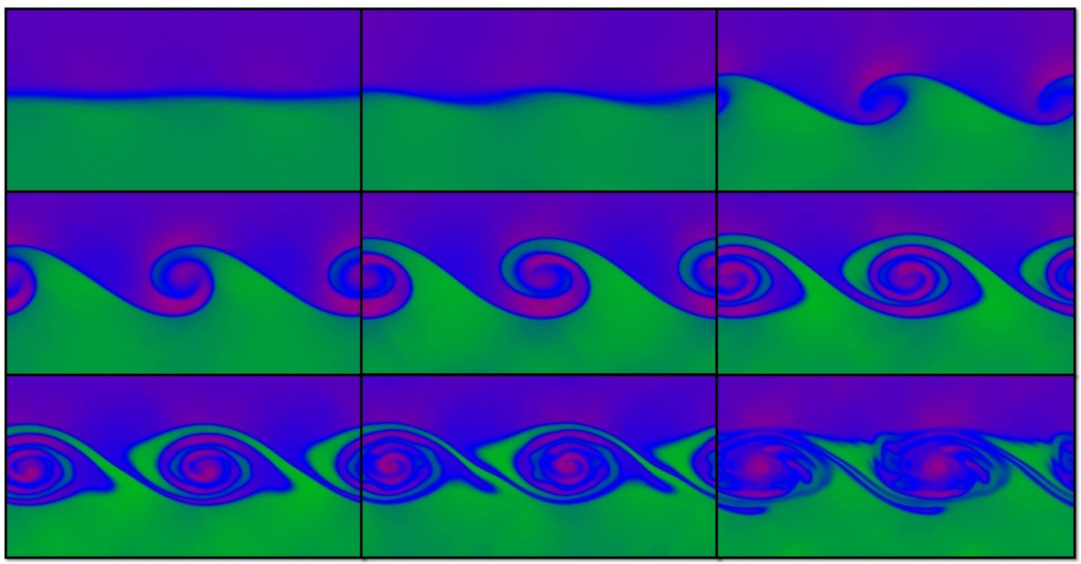

**Figura 2 - Evolução temporal da simulação com o Código Athena.**

A velocidade é definida como uma função senoidal, que cria a perturbação inicial. Veja na Figura 2 a evolução da simulação usando o código Athena.

A pertubação inicial, gerada pela diferença entre as velocidades e a diferença de densidade entre os dois fluidos, cria ocilações no fluido de maior densidade que dará origem aos vórtices. Ao longo do processo de interação, faz-se com que o vórtice se enrole e a mistura entre os dois fluidos se torne cada vez mais intensa.

## 4. CONCLUSÕES

O objetivo deste trabalho foi estudar a instabilidade de K-H usando o código Athena. Usando dois fluidos de densidade diferentes e velocidades com sentidos contrários. Com a simulação, conseguimos reproduzir, com clareza, os estágios em que foram encontradas nos objetos astrofísicos. Com o passar do tempo, na simulação, os gases se fundem.

## REFERÊNCIAS